\documentclass[nofootinbib,a4paper,pre,superscriptaddress,twocolumn,floatfix]{revtex4-1}  
\usepackage{times}
\usepackage{amsmath}
\usepackage{amsfonts}
\usepackage[mathcal]{euscript}
\usepackage{mathrsfs}
\usepackage{float}
\usepackage{grffile}
\usepackage{graphicx}
\usepackage{hyperref}

\newcommand{\h}{\Delta}

\newcommand{\uhp}{\mathbb{H}}
\newcommand{\D}{\mathbb{D}}

\begin{document}

\title{Domain walls and Schramm-Loewner evolution in the random-field Ising model}
\date{\today}
\author{Jacob D. Stevenson}
\author{Martin Weigel}
\affiliation{ Institut f\"ur Physik, Johannes Gutenberg-Universit\"at Mainz,
Staudinger Weg 7, 55128 Mainz, Germany }


\begin{abstract}
  The concept of Schramm-Loewner evolution provides a unified description of
  domain boundaries of many lattice spin systems in two dimensions, possibly
  even including systems with quenched disorder. Here, we study domain walls in
  the random-field Ising model. Although, in two dimensions, this system does
  not show an ordering transition to a ferromagnetic state, in the presence of
  a uniform external field spin domains percolate beyond a critical field
  strength. Using exact ground state calculations for very large systems, we
  examine ground state domain walls near this percolation transition finding
  strong evidence that they are conformally invariant and satisfy the domain
  Markov property, implying compatibility with Schramm-Loewner evolution
  (SLE$_{\kappa}$) with parameter $\kappa = 6$. These results might pave the
  way for new field-theoretic treatments of systems with quenched disorder.
\end{abstract}

\maketitle


\bibliographystyle{naturemagurl}

In the past decades, analytic techniques such as conformal field theory (CFT) and
Coulomb gas methods have led to a rather comprehensive understanding of critical
phenomena in two dimensions (2D). In particular, CFT allows for a complete
classification of 2D critical points, the exact determination of critical exponents
and, in some cases, even scaling amplitudes \cite{henkel:book}. This success is tied
to the fact that the conformal group is infinite-dimensional, however, which is true
only in 2D, and few of the results generalize to higher dimensions
\cite{weigel:99a}. Another difficulty for this approach arises for the important
class of systems with quenched disorder, such as diluted magnets, random-field
systems and spin glasses \cite{young:book}, since the non-unitary CFTs that are
believed to describe systems with quenched disorder are poorly understood
\cite{cardy:03b}.

While some geometrical aspects of critical phenomena had been previously worked out
using concepts from the Coulomb gas \cite{nienhuis:domb} and two-dimensional quantum
gravity \cite{duplantier:03}, a breakthrough was achieved with the description of
domain boundaries in terms of random curves in the plane in a framework dubbed
Schramm-Loewner evolution (SLE) \cite{bauer.2006}. In SLE, stochastic curves in the
plane are constructed from one-dimensional Brownian motion, thus classifying a
statistical ensemble of curves with only one parameter, the diffusion constant
$\kappa$.  Characteristic interfaces in many physical systems have been shown (in
some cases rigorously) to satisfy SLE$_\kappa$.  These include percolation
($\kappa=6$), self avoiding walks ($\kappa = 4/3$), as well as spin cluster
boundaries ($\kappa = 3$) and Fortuin-Kasteleyn cluster boundaries ($\kappa=16/3$) in
the Ising model. In recent years, close connections between SLE and CFT, including
links between probabilistic properties of SLE curves and scaling operators in CFT, or
between the central charge $c$ of the CFT and the diffusion constant $\kappa$ have
been established \cite{bauer.2006}. A number of numerical studies have found
interfaces in disordered systems to be (partially) consistent with SLE, in particular
the 2D Ising spin glass \cite{amoruso.2006,bernard.2007}, the Potts model on
dynamical triangulations \cite{weigel.2005b}, the random bond Potts model
\cite{jacobsen.2009}, and the disordered solid-on-solid model
\cite{schwarz.2009}. Such findings and the close link between SLE and CFT spur the
hope of a more complete understanding of systems with quenched disorder from a
field-theoretic perspective.

\section{Random-field Ising model}

Here, we study domain walls in the random-field Ising model (RFIM) with Hamiltonian
\cite{nattermann:97}
\begin{equation} 
  \mathcal{H} = - J \sum_{\langle i,j\rangle} s_i s_j - \sum_i h_i s_i,
  \label{eq:hamiltonian}
\end{equation} 
where the spins $s_i = \pm 1$ are located on the sites of a square lattice
and interact ferromagnetically with nearest neighbors. The
local fields $h_i$ are quenched random variables which, for the time being, we take
as drawn from a Gaussian distribution
with mean $H$ and standard deviation $\h$.
Since only the ratio $J/\Delta$ is relevant, we take $J=1$.
Random field models have a large number of experimental realizations
which are of technological importance such as superfluid helium, liquid
crystals in silica aerogels, Bragg glasses in high-$T_c$ superconductors, amorphous
solids\cite{stevenson.2008a} and ferroelectric materials \cite{feldman:01}. It was shown by Imry and Ma
\cite{imry:75} that random fields destabilize the ferromagnetic order in dimensions
$d\le 2$. For the case of 2D, it was argued by Binder
\cite{binder:83} that ferromagnetic order occurs only up to a break-up length scale
$L_b \sim e^{A / \h^2}$, which grows with decreasing disorder $\Delta$,
and that the system remains paramagnetic at scales $L> L_b$.  
Later, Aizenman and Wehr \cite{aizenmann:89a} proved that for $d\le 2$ the system
indeed has a unique Gibbs state, precluding the existence of an ordering
transition. On the contrary, for $d\ge 3$, $L_b$ diverges at the thermodynamic
transition point, below which the system is ferromagnetic \cite{nattermann:97}. For
non-zero average fields $H$, on the other hand, even in 2D the size of spin clusters
diverges at a critical value $H_c = H_c(\Delta)$
\cite{seppala.1998,seppala.2001,kornyei:07,stevenson.2010b}. However, the weight of
these clusters is sub-extensive, such that the free energy remains analytic and no
thermodynamic phase transition occurs. This phenomenon bears many similarities to the
Kert{\'e}sz line in ferromagnets in the absence of disorder \cite{kertesz:89a}. It is
this transition at non-zero $H$ that we study in this Letter.

To investigate the properties of domain walls in the RFIM we numerically compute
exact ground states of samples of random-field realizations. Ground states can be
found in polynomial time via a mapping to a minimum-cut/maximum-flow problem
\cite{dauriac.1985}. We employ a fast algorithm based on the idea of ``augmenting
paths''\cite{kolmogorov.2004}, which allows us to find ground states of systems of
$10^7$ spins in about $6$ s, such that the maximum system sizes exceed those of
previous studies \cite{seppala.1998,seppala.2001,kornyei:07} by about an order of
magnitude. We use a variety of domain geometries, partially with fixed spins to
enforce the occurrence of domain walls.  The calculations reported here were carried
out at either $\h = 2.5$ and $H=0.01362 = H_c(\h) \pm 0.00007$ or at $\h = 1.7$ and
$H=5.08 \times 10^{-4} = H_c(\h) \pm 0.07 \times 10^{-4}$.  For both cases the
breakup length scale is only a few lattice spacings, much less than the system sizes
we look at.  Both the breakup length scale and the critical external field were
determined using recipes laid out in Ref.~\cite{seppala.2001}.

\section{Schramm Loewner evolution}
In the framework of Loewner evolution one imagines a random curve $\gamma_t$ in the
plane as being continuously grown in time $t$ in a random process. Instead of
studying this process directly, one considers the evolution of a family $g_{t}: \uhp
\setminus \gamma_{t} \to \uhp$ of conformal maps that take the complement of
$\gamma_t$ in the upper half plane $\uhp$ to $\uhp$. Under this map the curve
$\gamma_t$, which lies on the boundary of $\uhp \setminus \gamma_{t}$, is taken to
the boundary of $\uhp$, i.e.\ to the real line. Assuming standard normalization and
boundary conditions, it turns out that $g_{t}$ is completely determined by the
one-dimensional function $\xi_t$, which corresponds to the image under $g_t$ of the
tip of the growing curve on the real line, via the differential equation
\begin{equation}
  \frac{\partial g_t(z)}{\partial t} = \frac{2}{g_t(z) - \xi_t}.
  \label{eqn:SLE_map}
\end{equation}
It was shown by Schramm \cite{schramm.2000} that if the ensemble of curves $\gamma_t$ is
conformally invariant and satisfies the domain Markov property (to be discussed
below), the one-dimensional random process described by $\xi_t$ must be Brownian
motion with zero mean and variance $\kappa t$.
For such SLE curves, many stochastic properties can be calculated
rigorously including, for instance, the fractal dimension $d_f = 1+\kappa/8$ or the
probability $P_\mathrm{LP}(x,y)$ that the curve $\gamma_t$ passes to the left of the
point $(x,y)$. The latter was proven by Schramm \cite{schramm.2001} 
for curves starting at the origin of the upper half plane
$\uhp$ to be
\begin{equation}
  P_\mathrm{LP}^{\kappa}(x,y) = \frac{1}{2} + \frac{\Gamma(4/\kappa)}{\sqrt{\pi}\,\Gamma
    \left(\frac{8-\kappa}{2\kappa} \right) } \frac{x}{y}
  {_2F_1}\left(\frac{1}{2},\frac{4}{\kappa};\frac{3}{2};
    -\left(\frac{x}{y}\right)^2 \right)
  \label{eq:schramm}
\end{equation}
where ${_2F_1}$ is Gauss' hypergeometric function.

\begin{figure}[t]
  \centering
  \includegraphics[width=\linewidth]{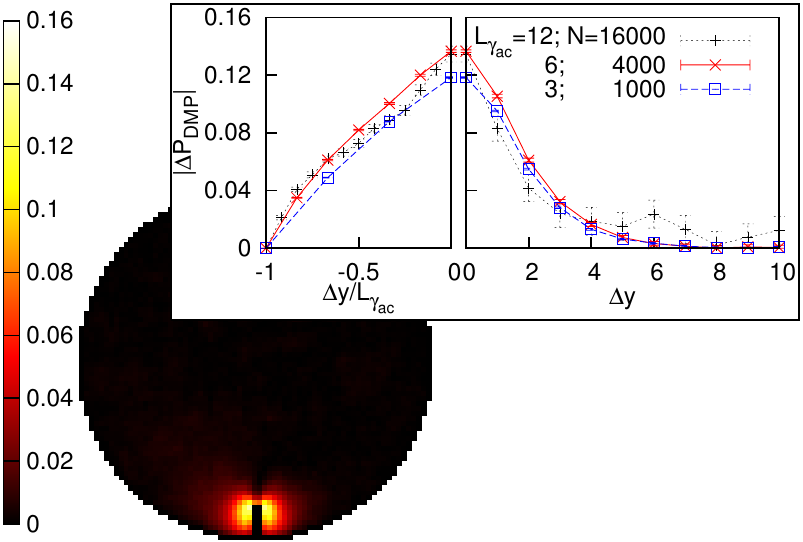}
  \caption{(Color online) The main panel shows the spatial distribution of 
    $|\Delta P_{\mathrm{DMP}}(x,y)|$ for a system of $4000$ spins in a circular domain
    and a vertical cut $\gamma_{ac}$ of length 6 starting at the bottom of the domain.
    Lighter colors correspond to larger deviations from the DMP.
    The right inset panel shows the decay of $|\Delta P_{\mathrm{DMP}}(35,y)|$
    with distance in the positive vertical direction $\Delta y$ (in lattice spacings) from 
    the point of
    maximum deviation.  Three
    different system sizes and correspondingly scaled cut lengths $\gamma_{ac}$
    are shown.
    The left inset panel shows the decay in the negative vertical direction (i.e. along 
    the cut) vs.\ $\Delta y$ scaled by the cut length.
    The tests were performed at $\h = 2.5$ and $H=1.362 \times 10^{-2} \approx
    H_c(\h)$.
    }
  \label{fig:DMP}
\end{figure}

\section{Domain Markov property}
The domain Markov property (DMP) formalizes the notion that a growing path of
the type described above is agnostic about its past. 
Let $P_{\D}(\gamma_{ab})$ be the probability measure of curves
$\gamma_{ab}$ in a domain $\D$ running between points $a$ and $b$ on the boundary of
$\D$, and let $c$ be a point in the interior of $\D$. Then, the DMP states that
\begin{equation}
  P_{\D}(\gamma_{cb} | \gamma_{ac}) = P_{\D \setminus \gamma_{ac} } (\gamma_{cb} ),
\label{eqn:DMP}
\end{equation}
i.e., the probability of $\gamma_{cb}$ is independent of whether $\gamma_{ac}$ is
preconditioned in domain $\D$, or whether $\gamma_{ac}$ is excluded from the domain
itself. While it is not debated that the DMP holds for domain boundaries in pure
lattice systems \cite{bauer.2006}, even off criticality, it has been argued that it
is likely not to survive the average over quenched disorder \cite{bernard.2007}.  For
the RFIM, we have checked the DMP numerically using the left passage probabilities
$P_\mathrm{LP}(x,y)$ instead of calculating the probability of all possible curve
segments $\gamma_{cb}$. For the l.h.s.\ of Eq.~(\ref{eqn:DMP}), this amounts to
picking out those configurations of the random fields that yield an interface along
$\gamma_{ac}$, while for the r.h.s., the interface is asserted to run along
$\gamma_{ac}$, for instance by fixing the corresponding spins with large magnetic
fields.  If the DMP holds, then
\begin{equation}
  \begin{split}
  \Delta P_{\mathrm{DMP}}(x,y) =& 
  \sum_{\gamma_{cb}} 
  P_{\D}(\gamma_{cb} | \gamma_{ac}) 
  P_{\mathrm{LP}}(x,y; \gamma_{cb} ) \\
  -& 
  \sum_{\gamma_{cb}} 
  P_{\D \setminus \gamma_{ac} } (\gamma_{cb} )
  P_{\mathrm{LP}}(x,y; \gamma_{cb} )
  \end{split}
\label{eqn:DMP_LP}
\end{equation}
will be identically zero.  We have studied system sizes of 1000, 4000, and 16\,000
spins with proportionally scaled cuts $\gamma_{ac}$ of length 3, 6, and 12,
respectively. For the largest system, we looked at $3\times 10^8$ ground state
configurations, of which only about $2800$ satisfied the conditioning along
$\gamma_{ac}$. As is clearly seen from $|\Delta P_{\mathrm{DMP}}(x,y)|$ shown in the
main panel of Fig.~\ref{fig:DMP}, the DMP does not survive the disorder average
exactly. The deviations are maximal around the tip of $\gamma_{ac}$ and fall off
rapidly with distance from $\gamma_{ac}$. As shown in the right inset panel the decay
of $\Delta P_{\mathrm{DMP}}$ with the vertical offset $\Delta y$ from the tip of
$\gamma_{ac}$ is nearly independent of the system size and the length of
$\gamma_{ac}$. In contrast, as shown in the left inset, the decay rate for
$\Delta y < 0$, i.e., along the cut, is proportional to the cut length. Perpendicular
to the cut, the decay rate (not shown) is again independent of system size. Hence,
the intrusion of deviations into the interior of the domain extends to only a few
lattice spacings and is largely independent of system size and cut length, such that
the DMP will be recovered in the scaling limit.  We find that the agreement with the
DMP in the scaling limit also holds off the critical percolation line.

\section{Left passage probability}
\begin{figure}[t]
\centering
\includegraphics[width=\linewidth]{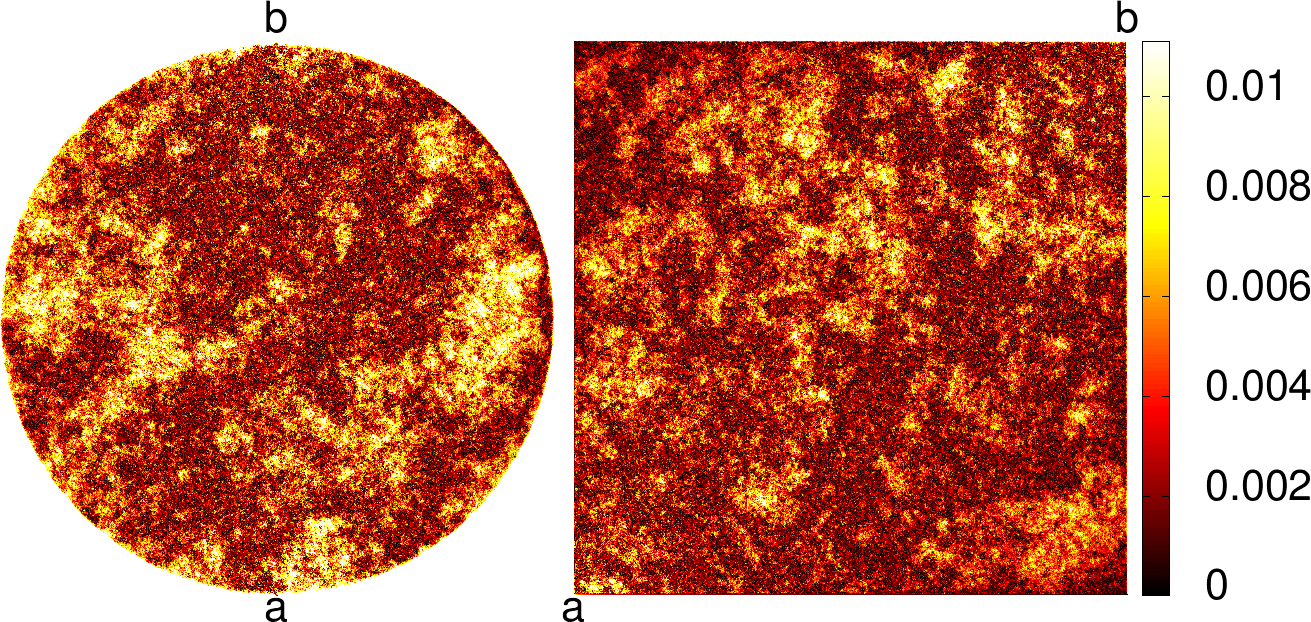}
\includegraphics{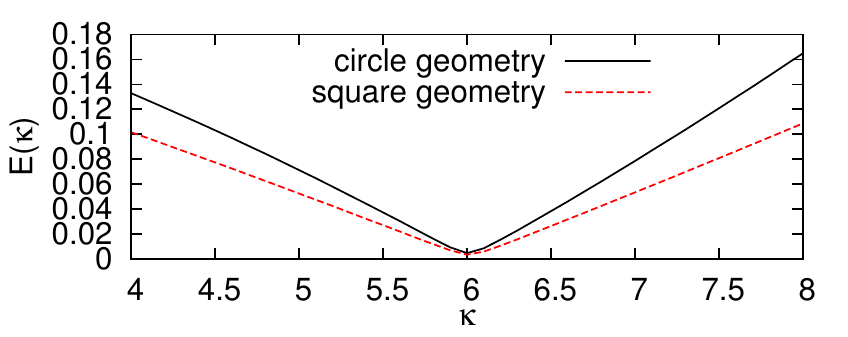}
\caption{Magnitude of deviation of the left passage probability for spin domain interfaces
  from the exact result of Eq.~\eqref{eq:schramm} for $\kappa = 6$ in circular and
  square domains. 
  The interfaces are constrained to run between points $a$ and $b$ as shown.
  The lower panel displays the spatially averaged deviation $E(\kappa)$
  as a function of the diffusion constant $\kappa$, showing a
  clear minimum close to $\kappa = 6$. 
  Calculations were performed for $10\,000$ disorder realizations on systems of $6 \times 10^6$ spins
  at $\h = 2.5$ and $H=1.362 \times 10^{-2} \approx H_c(\h) $.
}
\label{fig:left_passage}
\end{figure}

\begin{figure}
  \includegraphics{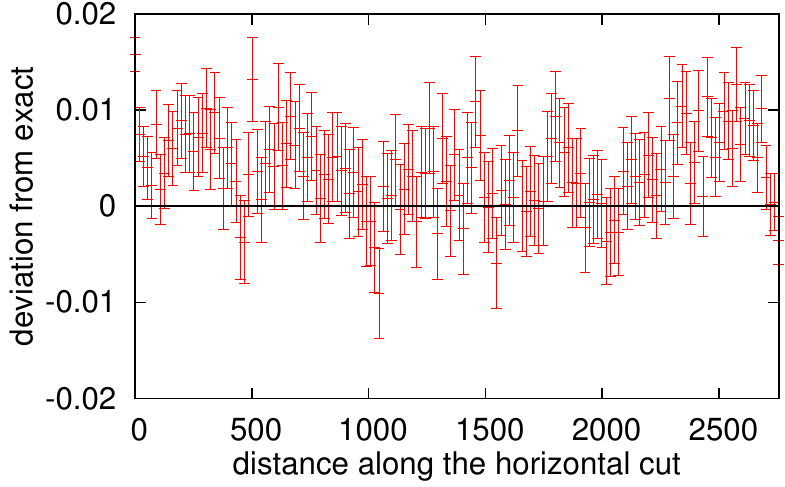}
  \caption{ Deviation from the exact left passage probability along
  a horizontal line crossing the center of the circle geometry that is shown in
  Fig.~\protect{\ref{fig:left_passage}.} 
  }
  \label{fig:left_passage_errorbars}
\end{figure}

We examined the agreement of the RFIM interfaces with the SLE expectations for the left
passage probabilities.  As the rigorous result of Eq.~(\ref{eq:schramm}) is
valid on the upper half plane $\uhp$, we performed our ground state calculations on
lattices embedded in domains 
$\D$ which have simple, closed-form conformal maps $w(z)$ to $\uhp$ \cite{schwarz.2009}.  By
fixing the boundary spins through the respective random fields, the interface was
forced to run between points $a$ and $b$ in $\D$ which are mapped to the origin and
infinity in $\uhp$, respectively. Numerical checks were performed for the unit circle
with the interface between $-i$ and $i$, which is mapped to $\uhp$ by $w(z) = i
\frac{1+z}{1-z}$, as well as the unit square with interfaces defined from $0$ to
$1+i$, which is mapped to $\uhp$ by $w(z) = - \wp ( 1+i-z; 1,i)$ where
$\wp(z;w_1,w_2)$ is the Weierstrass p-function. 
Looking at the left passage probability for multiple domains additionally acts as a check of 
conformal invariance.
For a
quantitative comparison we considered the mean square deviation of the computed
left passage probability $P_\mathrm{LP}$ from the exact result $P_\mathrm{LP}^\kappa$
of Eq.~(\ref{eq:schramm}),
\begin{equation}
  E(\kappa) =  \left< \left[ P_\mathrm{LP}(x,y) -
        P_\mathrm{LP}^{\kappa}(x,y) \right]^2 \right>_{\D}^{1/2},
      \end{equation}
where $\langle \cdot \rangle_\D$ denotes a spatial average over $\D$, excluding the
vicinity of the fixed boundary spins. This quantity is shown in
Fig.~\ref{fig:left_passage} as a function of $\kappa$ for the circle and square
geometries. In both cases $E(\kappa)$ is minimal for $\kappa$ within $0.05$ of the
value $\kappa=6$. The spatial dependence of the deviation from
$P_\mathrm{LP}^{\kappa=6}(x,y)$ is also shown in Fig.~\ref{fig:left_passage}
and, for a horizontal cut through the domain, in
Fig.~\ref{fig:left_passage_errorbars}. There appears to be no systematic
deviation.

\section{Crossing probability}
\begin{figure}[t]
\centering
\includegraphics{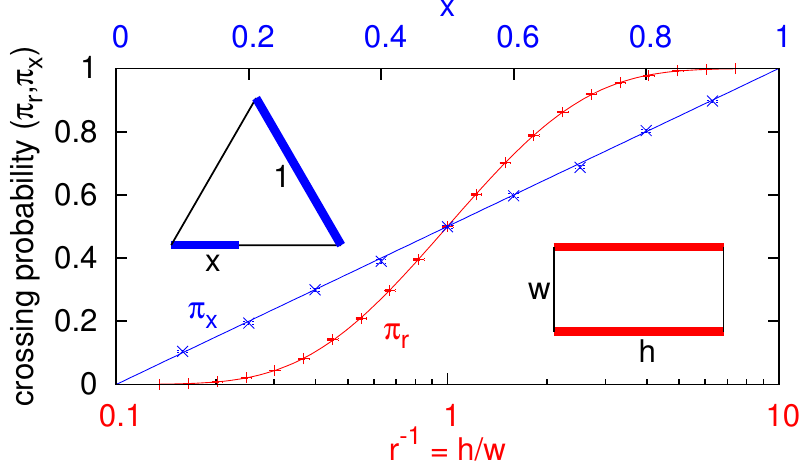}
\includegraphics{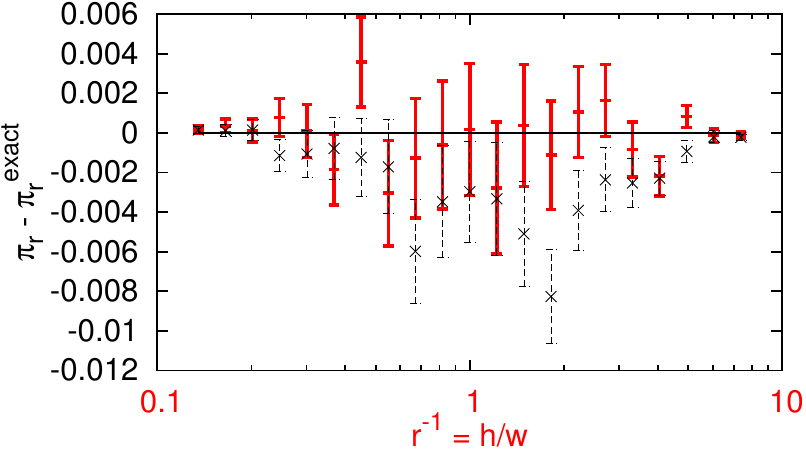}
\caption{
  Upper panel: Crossing probabilities for spin clusters in the 2D RFIM in rectangular
  (lower scale) and triangular (upper scale) domains.  Calculations were performed at
  $\h=1.7$ and $H = 5.08\times 10^{-4} \approx H_c(\h)$ for 20--30$\times10^3$
  disorder realizations for the rectangular domain and 7$\times10^3$ realizations
  for the triangular domain. The solid lines indicate the exact results
  for percolation clusters (SLE$_{\kappa = 6}$) derived in
  Refs.~\cite{cardy.1992,smirnov.2001}. Lower panel: Deviations of the numerical
  results from the exact expression for percolation for the rectangular domain.
  Deviations in the triangular domain are similar. Shown with dashed error bars are
  data calculated at $\h = 1.7$, but at an external field $H = 4.71\times 10^{-4}$.
  Some systematic deviations from SLE expectations are already seen for this slight
  detuning from criticality.
}
\label{fig:crossing}
\end{figure}

From the presented results it is evident that spin domain interfaces in the 2D RFIM
satisfy the domain Markov property and are conformally invariant at the percolation
threshold $H=H_c(\h)$ in the scaling limit, and thus are described by SLE.
Furthermore, the parameter $\kappa$ appears to be consistent with $\kappa = 6$ to
high precision. To corroborate these findings, we tested our results for
compatibility with the exact formulas for the crossing probabilities of percolation
clusters, another system with $\kappa = 6$ \cite{cardy.1992}. These give the
probability of finding a cluster of a given species (say up spins) which touches two
non-adjacent segments of the boundary of a domain. In particular, the probability
$\pi_r$ of a domain touching both the top and bottom of a rectangle of aspect ratio
$r$ is known to depend only on $r$ \cite{cardy.1992}.  Similarly, in a domain defined
by an equilateral triangle, the probability of a cluster crossing from a fraction $x$
of one boundary edge to the opposite edge is $\pi_x = x$ \cite{smirnov.2001} (see
Fig.~\ref{fig:crossing} for a schematic representation). Figure \ref{fig:crossing}
shows these exact percolation results together with numerical simulation data for the
RFIM at $\h = 1.7$ and $H=5.08 \times 10^{-4}\approx H_c(\h=1.7)$ for a system of
$6\times 10^6$ spins for both the rectangle and triangle geometries. We find very
good agreement with the percolation results, cf.\ the lower panel of
Fig.~\ref{fig:crossing}.  We also show results calculated with an external field
$H=4.71 \times 10^{-4} \approx 0.93\,H_c$. Systematic deviations from SLE can be
clearly seen, even for this slight detuning from criticality, indicating the high
sensitivity of our tests.  

\section{Fractal dimension}
\begin{figure}
  \includegraphics{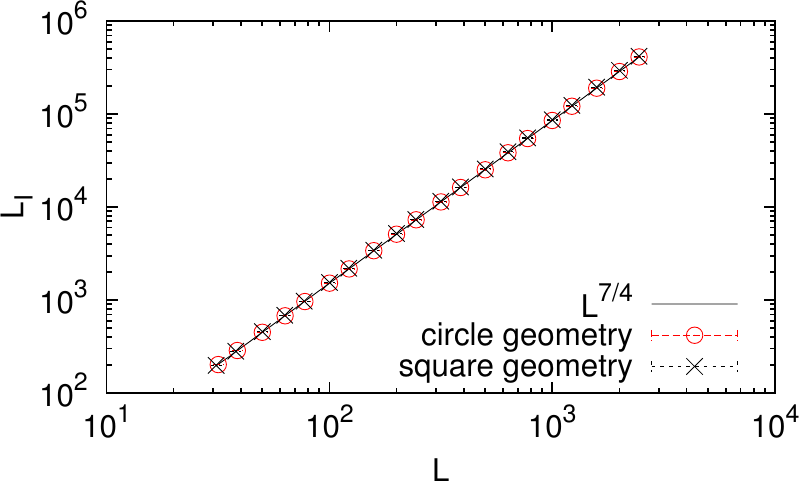}
  \caption{ Scaling of interface lengths with system size.  The best fit
  values --- $d_f = 1.7506(9)$ for the circle geometry and  $d_f = 1.7514(14)$ for
  the square geometry --- are in excellent agreement with the value expected for SLE$_6$, $d_f=7/4$. 
  The calculations were carried out at $\h=2.5$ and $H=1.362 \times 10^{-2} \approx H_c(\h)$.
  }
  \label{fig:fractal_dimension}
\end{figure}

One of the rigorous results for curves described by $\mathrm{SLE}_\kappa$ is their
fractal dimension, which is given, for $\kappa \le 8$, by $d_f = 1+\kappa/8$.
We numerically determined the fractal dimension for a range of
different geometries and display the results in Fig.~\ref{fig:fractal_dimension}. We find that corrections to scaling for the
interface length
$L_I$ are well described by the form
\begin{equation}
  \label{eq:fractal_corrected}
  L_{I} = a L^{d_f} (1+b/L).
\end{equation}
We take $L$ as the square root of the number of spins for the case of non-rectangular
domains.  For the circle geometry we find $d_f = 1.7506(9)$, while for the square
geometry we arrive at $d_f = 1.7514(14)$, using $\Delta = 2.5$ and $H=1.362
  \times 10^{-2} \approx H_c(\h)$ in both cases. These results are perfectly
compatible with $d_f = 7/4 = 1.75$ expected for SLE curves with $\kappa = 6$.

\section{Brownian motion}
As the most direct test for SLE, we 
studied the one-dimensional stochastic process (or driving function)
$\xi_t$ generated by the Loewner map $g_t$ according to Eq.~\eqref{eqn:SLE_map} as
applied to domain walls in the RFIM. For a lattice system, the family of maps $g_t$
is realized as a discrete series of maps $g_{i}$ iteratively removing a small section
from the beginning of the curve. For this purpose, $g_{i}$ is approximated using a
vertical slit map \cite{kennedy.2008}
\begin{equation}
  g_i(z) = i \sqrt{ -(z-\xi_i)^2 - 4 \Delta t_i } + \xi_i.
  \label{eqn:slit_map}
\end{equation}
Here, $\xi_i$ and $\Delta t_i$ are determined through $\xi_i = x_{i,i-1}$ and $\Delta
t_i = y_{i,i-1}^2/4$, where $x_{i,i-1}$ and $y_{i,i-1}$ are the coordinates of the
$i$'th segment of the curve after undergoing the $i-1$ successive maps $g_{i-1}\circ
\ldots \circ g_{1}$. The parameter $\xi_i$ is the value of the driving function
$\xi_t$ sampled at time $t_i = \sum_{j\le i} \Delta t_j$.  The complex square root in
Eq.~\eqref{eqn:slit_map} is calculated, as usual, with the branch cut along the
negative real axis.
We studied the statistics of $10\,000$ interfaces generated in a half disc, optimally
mimicking the full space $\uhp$. The interface is initiated at the origin by two
fixed spins and is considered ended when it touches the curved boundary. We used
systems of 6 million spins at $\h = 2.5$ and $H=1.362 \times 10^{-2} \approx H_c(\h)$.
We find that the variance of the driving function calculated from the interfaces is
$\hat{\kappa} = \langle (\xi_t - \langle \xi_t \rangle )^2 \rangle/t = 6.086 (87)$,
and the normalized mean is $\hat{\xi} = \langle \xi_t \rangle / \sqrt{ \hat{\kappa} t
} = 0.017 (10)$, perfectly compatible with $\mathrm{SLE}_{\kappa=6}$. Using a
Kolmogorov-Smirnov test \cite{kennedy.2008,james.2006}, we further checked that the $\xi_t$
are normally distributed and find a $p$-value of $p=0.17$, indicating consistency
with a normal distribution. To check for the statistical independence of the
increments of $\xi_t$, we divided $\xi_t$ into $n$ increments evenly spaced in time
and checked whether the signs of these increments follow a $\chi^2$ distribution with
$2^n-1$ degrees of freedom \cite{kennedy.2008}. For $n=10$, we find a $p$-value of
$p=0.18$ indicating consistency with the assumption of statistical independence.  As
a further check of conformal invariance, we performed the same tests on interfaces
originating in the two domains of Fig.~\ref{fig:left_passage}, and found similar
agreement with SLE.

\begin{figure}[t]
\centering
\includegraphics[width=\linewidth]{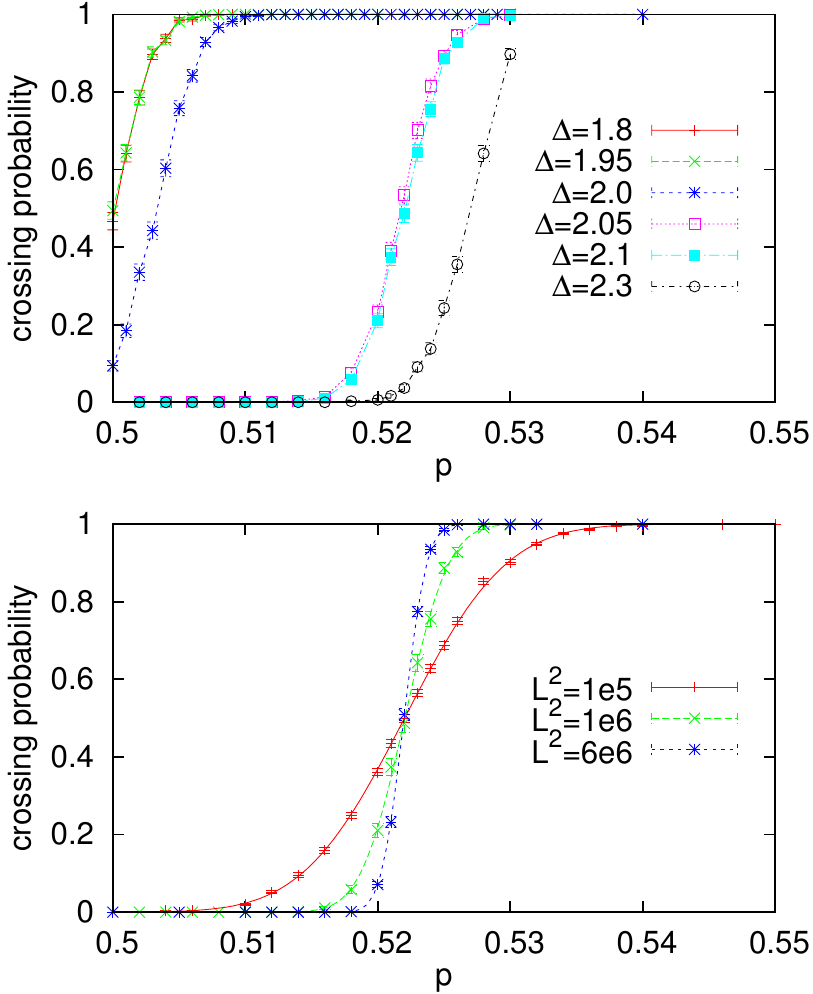}
\caption{
Crossing probabilities for the binary distribution of fields as a function of
the fraction $p$ of up fields.  The upper panel shows the dependence of these
curves on disorder strength (for system size $L^2=10^6$).  The bottom panel shows the crossing of these
curves for different system sizes at the critical point $p_c$ (at disorder strength $\h = 2.1$).
}
\label{fig:pc_fig}
\end{figure}

\begin{figure}[t]
\centering
\includegraphics[width=\linewidth]{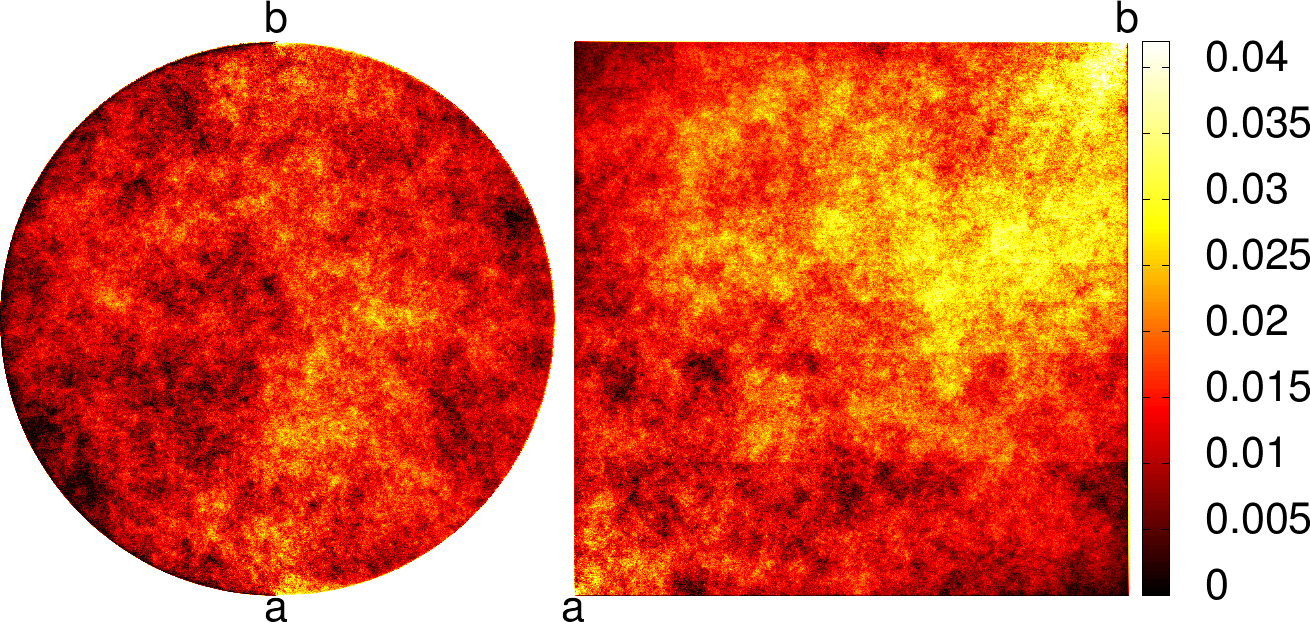}
\includegraphics{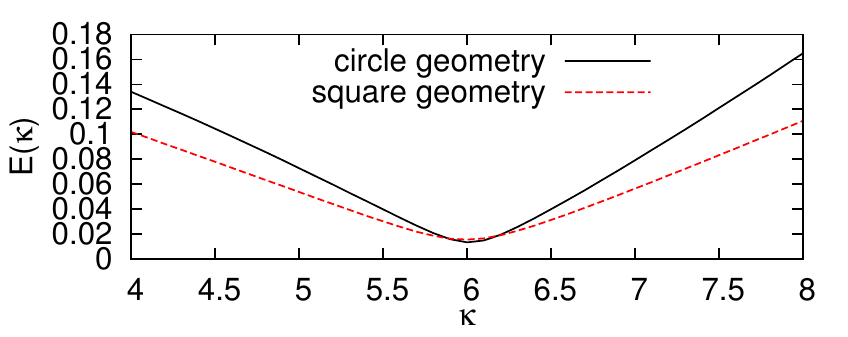}
\caption{
Left passage probabilities for the binary distribution of fields.
See Fig.~\ref{fig:left_passage} for a detailed description of the setup.
The calculations were carried out at $\h = 2.1$ and $p = 0.522 \approx p_c(\h)$
for systems of 6 million spins.  Averages were performed over 
$10$ randomly chosen ground states for each of 
$6000$ disorder configurations.
}
\label{fig:left_passage_binary}
\end{figure}

\section{Binary fields}
Finally, we also considered a binary (Bernoulli) field distribution, where each local random
field $h_i$ takes on the value $\h$ with probability $p$ and $-\h$ with probability $1-p$.
Because of the discrete nature of the distribution, this system has a massive
ground-state degeneracy, and behaves rather differently from
the Gaussian RFIM, at least at zero temperature. Although polynomial algorithms for
enumerating all ground states are known \cite{bastea.1998b}, handling
all ground states becomes impractical for larger system sizes. Here, instead, we
sample from the ground-state manifold by adding a tiny noise term (normally
distributed with strength $\delta$) to the Bernoulli field distribution. For
sufficiently small $\delta$, the resulting state is also a ground state of the
noise-less system, and, importantly, is selected without bias from among the
degenerate ground states. We find that there exists a geometric transition where spin
clusters diverge at $p=p_c(\Delta)$, in analogy with $H_c(\Delta)$ for the Gaussian
case.  
This is illustrated in Fig.~\ref{fig:pc_fig}, where we have plotted, for a number of different disorder strengths, the probability
of finding a spin-up cluster that touches both the top and bottom boundaries of a square
geometry.  Shown in the bottom panel is the size dependence of these curves, demonstrating
that different sizes cross at $p_c$.  We also point out that this crossing of curves happens when the crossing probability is $0.5$ as expected.
Interestingly, as can be seen in in the upper panel of Fig.~\ref{fig:pc_fig}, for $\Delta < 2$ this transition appears to occur at the
constant value $p_c(\Delta < 2)=1/2$, while $p_c(\Delta \geq 2) > 1/2$.  We test
agreement with SLE predictions by looking at the fractal dimension at $p_c$, finding $d_f =
1.746(2)$, and by looking at the left passage probability, the results of which are shown in
Fig.~\ref{fig:left_passage_binary}.  Both properties are consistent with SLE for
$\kappa = 6$.

\section{Conclusions}
We have studied the properties of spin cluster interfaces in the ground state of the
Gaussian random field Ising model at values of the external field strength $H$ where
the size of the clusters diverges. For this $T=0$ system with quenched disorder, the
domain Markov property was shown to be satisfied in the scaling limit.  Together with the
conformal invariance of the interfaces deduced from Schramm's formula and the
crossing probabilities, it is shown clearly that the spin domain interfaces satisfy
$\mathrm{SLE}_{\kappa=6}$, corresponding to pure percolation.
The fractal dimension is in
perfect agreement with these observations, contrary to the case of the solid-on-solid
model studied in Ref.~\cite{schwarz.2009}, where $\kappa\approx 4$ was found from
Schramm's formula, but $d_f \approx 1.25 \ne 1+\kappa/8$. Studying the SLE map
directly, we have shown that the driving function describes Brownian motion. 
The consistency with SLE carries over to the case of binary random fields, where degeneracies
occur. This is in contrast to the observations for the spin glass model, where domain
walls appear to be only described by SLE for continuous disorder distributions
\cite{amoruso.2006,bernard.2007,gusman:08}. The 2D RFIM thus seems to provide a
paradigmatic example where SLE is realized in all known aspects in a system with
quenched disorder, nourishing the hope for a more systematic treatment of systems
with quenched disorder in field theory.

The authors acknowledge computer time provided by NIC J\"ulich under grant No.\ hmz18
and funding by the DFG through the Emmy Noether Program under contract
No. WE4425/1-1.


\end{document}